\documentclass[preprint2]{aastex}
\slugcomment{Submitted to Astrophysical Journal Letters}
\lefthead{Martinez and Shlosman}
\righthead{Buckling Stellar Bars}

\def \m3{{\rm Mark III}}

\def\gtorder{\mathrel{\raise.3ex\hbox{$>$}\mkern-14mu
    \lower0.6ex\hbox{$\sim$}}}
\def\ltorder{\mathrel{\raise.3ex\hbox{$<$}\mkern-14mu
    \lower0.6ex\hbox{$\sim$}}}

\begin{document}
\lefthead{Martinez and Shlosman}
\righthead{}

\title{WHY BUCKLING STELLAR BARS WEAKEN\\ IN DISK GALAXIES}

\author{Inma Martinez-Valpuesta\altaffilmark{1} and Isaac Shlosman} 
\affil{Department of Physics and Astronomy,
University of Kentucky, Lexington, KY 40506-0055, USA\\
email: {\tt martinez@pa.uky.edu,} {\tt shlosman@pa.uky.edu}}

\altaffiltext{1}{Also: at Department of Physics, Astronomy \& Mathematics,
University of Hertfordshire, Herts AL10 9AB, UK}

\begin{abstract}
Young stellar bars in disk galaxies experience a vertical buckling instability
which terminates their growth and thickens them, resulting in a characteristic
peanut/boxy shape when viewed edge on. Using $N$-body simulations of galactic
disks embedded in live halos, we have analyzed the bar structure throughout
this instability and found that the outer (approximately) third of the bar
dissolves completely while the inner part (within the vertical inner Lindblad
resonance) becomes less oval. The bar acquires the frequently observed
peanut/boxy-shaped isophotes. We also find that the bar buckling is
responsible for a mass injection above the plane, which is subsequently
trapped by specific 3-D families of periodic orbits of  particular shapes
explaining the observed isophotes, in line with previous work. Using a 3-D
orbit analysis and surfaces of sections, we infer that the outer part of the
bar is dissolved by a rapidly widening stochastic region around its corotation
radius --- a process related to the bar growth. This leads to a dramatic
decrease in the bar size, decrease in the overall bar strength and a mild
increase in its pattern speed, but is not expected to lead to a complete bar
dissolution. The buckling instability appears primarily responsible for
shortening the secular diffusion timescale to a dynamical one when building
the boxy isophotes. The sufficiently long timescale of described evolution,
$\sim 1$~Gyr, can affect the observed bar fraction in local universe and at
higher redshifts, both through reduced bar strength and the absence of dust
offset lanes in the bar. 
\end{abstract}


\section{Introduction}

Stellar bars are among the extreme signatures of a breakup of axial
symmetry in galactic disks. As such, they serve as an impetus for secular and
dynamical evolution of galaxies at all redshifts. The bar formation and growth
largely depend on the
efficiency of angular momentum redistribution, i.e., the ability of the inner
(bar unstable) disk to lose the angular momentum and of the outer disks, halos
and interactions to absorb it (Athanassoula 2003). The growth of {\it
numerical} collisionless bars can be characterized by increase  in the
amplitude of $m=2$ mode and their prolateness, i.e., decrease of equatorial
axial ratio $b/a$. Bars in this initial stage of evolution appear to be as flat
as the disk of their origin, with shortest-to-longest axis ratio $c/a\sim
0.1$. This is supported by numerical modeling of bar instability over
nearly three decades (e.g., Sellwood \& Wilkinson 1993).

Initial growth of model bars is terminated by the so-called vertical buckling
instability, first detected in numerical simulations of Combes \&
Sanders (1981) and given two alternative explanations: a resonant bending
(Combes et al. 1990; Pfenniger \& Friedli 1991) and
firehose instability (Raha et al. [1991], who invoked the explanation by
Toomre [1966]; Merritt \& Sellwood 1994). This instability leads to the
vertical thickening of the bar
on a dynamical timescale via spectacular breakup of vertical symmetry.
Interest to this phenomenon has been further amplified by the similarity
between frequently observed peanut- and boxy-shaped bulges in edge-on disk
galaxies (e.g., Burbidge \& Burbidge 1959; Jarvis 1986; Shaw 1987; Bureau \&
Freeman 1999; Merrifield \& Kuijken 1999) and those obtained in numerical
models (Patsis, Skokos \& Athanassoula 2002a; Aronica et al. 2003; O'Neil
\& Dubinski 2003). The addition of an {\it inhomogeneous} dissipative component
to the disk acts as to weaken the buckling and to wash out the boxy inner
shape, resulting in a more `classical' bulge with the shape parameter $n$
larger by a factor of 2 (Berentzen et al. 1998).
 
Furthermore, numerical simulations capturing this
evolution of stellar bars show them to weaken dramatically during
the buckling instability. However, the reason for this drop in the bar
strength was never explained. Is it caused by the buckling? Can it
lead to a complete bar dissolution (e.g., Raha et al. 1991), is the bar 
weakened temporarily or permanently? What fraction of the orbits and which 
orbits stop to support the bar potential?

In this Letter we focus on the physical reasons for this behavior and provide
quantitative answers to these questions. Our results are based on the
self-consistent 3-D $N$-body simulations and a subsequent
nonlinear 3-D orbit analysis of the time-dependent numerical models.

The emerging connection between the peanut/boxy bulges in edge-on disks and
the buckling instability allows, in principle, to deduce the face-on properties
of galaxies in their most unfavorable orientation. This instability appears to
be important in understanding dynamical and secular evolution of barred
galaxies. It bears direct consequences for radial redistribution of their
stellar and gasesous components, and can affect the distribution of star
formation sites --- both depend strongly on the bar strength and its pattern
speed. Lastly, this is one of the processes which contribute to the growth of
the pseudo-bulges (e.g., review by Kormendy \& Kennicutt 2004).

The nature of the buckling instability is being slowly understood
and recent progress is based on the analysis of the orbital structure of barred
(Pfenniger 1984; Skokos, Patsis \& Athanassoula 2002a,b) and unbarred (Patsis
et al. 2002b) disks and spheroidal components (Binney \& Petrou 1985; May et
al. 1985), and is related to the shapes of dominant orbital families.
Observationally, significance of peanut/boxy bulges follows directly from
their abundance --- almost half of all edge-on disks exhibit them (L\"utticke,
Dettmar \& Pohlen 2000). The emerging picture is that of two processes:
of the firehose instability leading to the buckling in the midplane
of the bar and of resonant heating, trapping the particles around stable
3-D orbits which furnish the bar with the boxy-shaped bulge. The energy
deposited initially in the characteristic wavelength of buckling instability
subsequently cascades down to smaller wavelengths, increasing the vertical
dispersion velocities within the region.
 
More specifically, Pfenniger \& Friedli (1991) have identified 3-D orbital
families,\footnote{Their projections onto the potential midplane are elongated
with the bar, similarly to the main orbit family supporting the bar} which if
populated will provide the bar with
the specific `butterfly' or peanut shape when viewed edge-on along the minor
axis. These families originate at the vertically-unstable gap of plane
periodic orbits, in other words, at the vertical inner Lindblad resonance
(VILR) which is almost always present in the bar, at roughly $1/3-2/3$
of its corotation radius. In this picture, the growth of the bar is limited by
formation of unstable plane orbits (Pfenniger 1984). Those diffuse across the
VILR, leading to the bar thickening in the resonance region.  

\section{Results}

We have used version FTM-4.4 of $N$-body code (e.g., Heller \&
Shlosman 1994) with $N=6\times 10^5$ collisionless particles, which represent
the stellar disk and dark halo components. In a
number of runs, the particles have been distributed initially according to Fall
\& Efstathiou (1980) analytical model which consists of an iteratively-relaxed
halo and an exponential disk. The dynamical time is $4.7\times 10^7$~yrs, and
$r$ and $z$ are the radial and vertical coordinates in the disk. The initial
conditions for the model described here are chosen such that the disk/halo
mass ratio within 10~kpc is unity. The halo has a flat density core of 2~kpc
to avoid excessive stochastic behavior associated with the central cusps
(El-Zant \& Shlosman 2002). The radial and vertical disk scalelengths are
2.85~kpc and 0.5~kpc and the Toomre's $Q$ parameter is 1.5. Gravitational
softening of 160~pc was used. For the orbital analysis, we use the updated
algorithm described in Heller \& Shlosman (1996). All the discussion involving
3-D orbital structure and resonances in the bar are based on this {\it
nonlinear} formalism and differ substantially from the epicyclic (linear)
approximation. The bar pattern speed, $\Omega_{\rm b}$, is calculated from the
phase angle of the $m=2$ mode. 
The energy and angular momentum in the system are conserved to within
approximately 1\% and 0.05\% accuracy, respectively. Our results appear to be
reasonably independent of $N$. 

\begin{figure}[ht!!!!!!]
\vbox to3.1in{\rule{0pt}{3.1in}}
\includegraphics{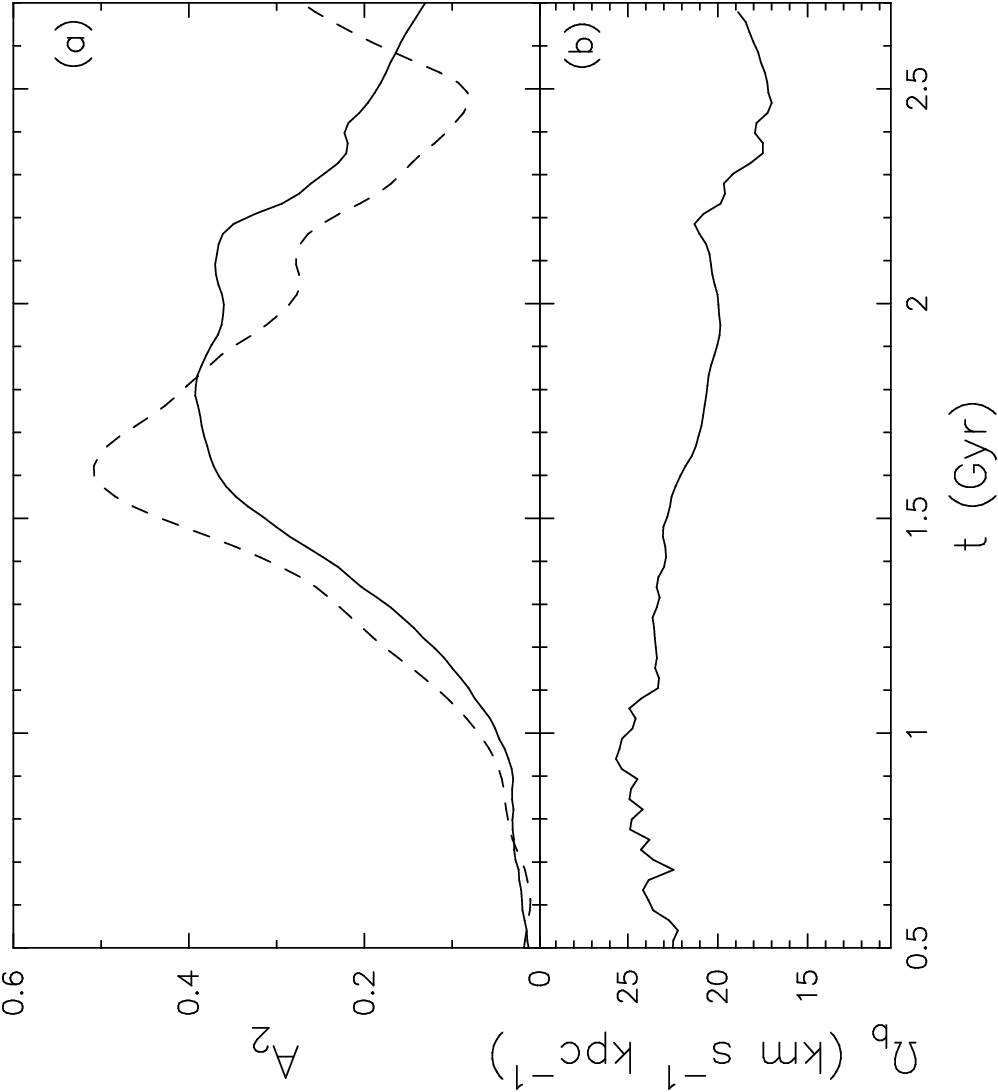}
\caption{{\it (a).} Evolution of bar $m=2$ amplitude, $A_2$, for $r=0-4$~kpc
(solid) and $r=6-10$~kpc (dashed) --- inner and outer bar parts; {\it (b).}
Bar pattern speed $\Omega_{\rm b}$.   
\label{fig:a2ampl}
}
\end{figure}

The initially  axisymmetric model developes a prominent bar in about three
rotations (as measured by $m=2$ amplitude $A_2$, Fig.~1a), which starts to
break against the halo and the outer disk, reducing its pattern speed
(Fig.~1b). Between about $t\sim 1.6-2.4$~Gyr, the bar becomes vertically
unstable and buckles, breaking the symmetry with respect to the disk
equatorial midplane
(Fig.~2). After $t\sim 2.4$~Gyr, the bar profile in the $rz$ plane again tends
towards symmetry, but the inner bar part has now a larger vertical thickness,
$c/a\sim 0.3$, and its isophotes have  acquired a boxy appearance, in
accordance with previous work on this subject. More careful analysis reveals
additional fundamental changes and transformations in the bar. This subtle bar
evolution can be followed through changes in its orbital structure. 

\begin{figure}[ht!!!!!!]
\vbox to3.2in{\rule{0pt}{3.2in}}
\includegraphics{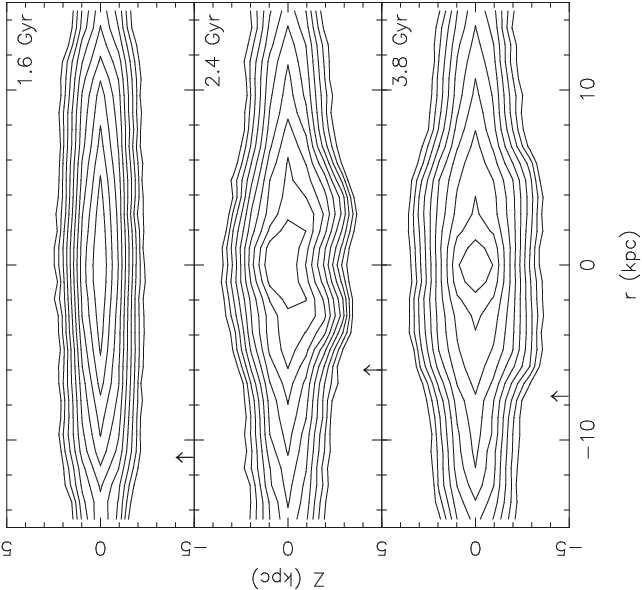}
\caption{From thin bar -- to buckled one -- to boxy bulge: snapshots in the
$rz$ plane at $t=1.6$~Gyr (upper), 2.4~Gyr (middle) and 3.8~Gyr (lower) given
by projected isodensities, whose values have been preserved from frame to
frame. The bar is oriented perpendicular to the line-of-sight. Note the
dramatic shortening of inner contours. The vertical arrows indicate the
approximate extent of the bar.
\label{fig:buckling}
}
\end{figure}

First, the bar size and strength change in a particular way
during the short period of buckling. The size of the bar is defined
here in the observational context (following Knapen, Shlosman \& Peletier
2000). Namely, the bar is ``detected'' by the maximal ellipticity of the
face-on fitted bar isophotes (i.e., isodensities), and their constant position
angle (PA). Its size is taken to be the maximal radius of PA$=
const.$\footnote{Note, that $A_2$ maximum is {\it not} a good measure of the
bar size as higher harmonics, $m=4$ and 8, have a substantial contribution} 
We have verified {\it aposteriori} that this definition does not contradict the
bar size obtained from the extent of the largest {\it stable} periodic orbit
supporting the bar figure.

The isodensity ellipse fitting to the bar at $t=1.6$~Gyr resulted in
rising ellipticity along the bar, from $\sim 0.45$ at the innermost to about 0.78
at $r\sim 8-9$~kpc and its subsequent drop. For $t=2.4$~Gyr, the bar
ellipticity stayed flat, $\sim 0.5$, up to $r\sim 5$~kpc and dropped sharply
for larger $r$. The inner part of young numerical bars hence appears by far
more non-axisymmetric than in their observational counterparts.

Prior to buckling, the bar grows and extends to nearly its corotation radius,
$r_{\rm CR}$, which increases with time due to the decreasing $\Omega_{\rm b}$.
Between $t\sim 1.9-2.4$~Gyr, the bar decreases in its length $r_{\rm b}$ by
$\sim 1/3$, but resumes its grows afterwards. The ratio $r_{\rm
CR}/r_{\rm b}$ is about 1.05 between $t\sim 1.4-1.9$~Gyr, increases to 1.7 at
$t\sim 2.4$~Gyr and drops to 1.4 thereafter. At $t\sim 2.4$~Gyr, the bar size
appears to correspond to $r_{\rm VILR}\sim 0.5-0.6 r_{\rm CR}$, or in other
words, to the radius of an unstable gap in the main family of $xy$ planar
orbits supporting the bar.

\begin{figure*}[ht!!!!!!!!!!]
\vbox to3.3in{\rule{0pt}{3.3in}}
\includegraphics{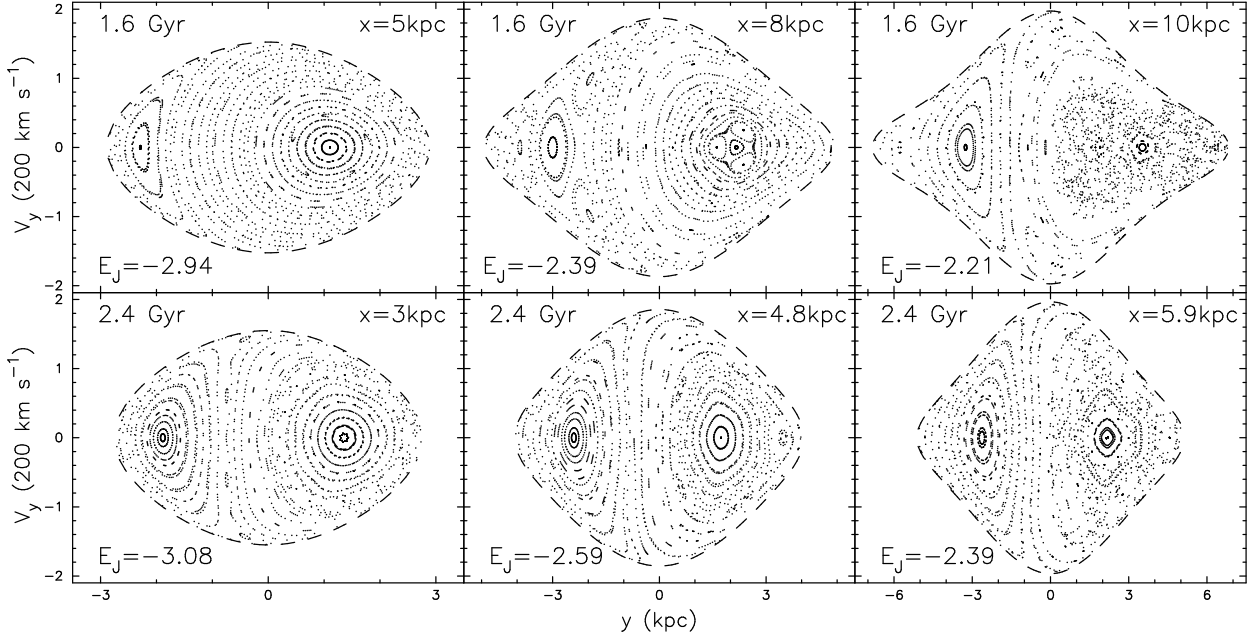}
\caption{Surface of section diagrams in the ($y, v_{\rm y}$) plane showing
stochastic orbit-dominated strong bar replaced by a shorter milder bar with
a larger fraction of regular orbits in the bar's midplane, $z=0$. This
evolution happens concurrently with the buckling instability. Regular orbits
form closed curves surrounding the fixed points of parent orbits. Here $x$ and
$y$ are oriented along the major and minor axes of the bar. In the figure,
$y<0$ side represents retrograde orbits, $y>0$ -- prograde ones. Jacobi energy
cuts, $E_{\rm J}$, corresponding to $r_{\rm b}$ (right column), $0.8r_{\rm b}$
(mid-column) and $0.5r_{\rm b}$ (left column) have been made at two different
times, $t=1.6$~Gyr (upper frames) and $t=2.4$~Gyr (lower frames), when the bar
size drops from $\sim 10$~kpc to below 6~kpc. Direct and retrograde orbits
dominate the phase space deeper inside the bar (distances to the center are
given in the right upper corners). Note two effects: {\it (i)} much smaller
fraction of the phase space occupied by regular orbits in the outer stronger
bar (upper right frame) compared to dissolved outer and overall weakened bar
(lower right frame); and {\it (ii)} the stochastic region expands inwards from
the corotation region of the bar, dissolving its outer part, outside $x\sim
5.9$~kpc.   
\label{fig:}
}
\end{figure*}
 
Second, as Fig.~1a reveals, the weakening of the bar starts in the outer part
and propagates inwards, and while the amplitude of the inner part drops by a
factor of $\sim 3-4$, the outer bar basically {\it dissolves} and its $A_2$
amplitude tends to zero, before it {\it rebounds} and grows again. 

Third, the mass within the central $r \sim 1$~kpc jumps almost by a factor of 2 
between $t=1.6-3.8$~Gyr, and by about 20\% within $r=3$~kpc, leading to a much
more centrally concentrated system. This mass concentration grows rather
`impulsively,' and can in principle be responsible for periods of a mild
spinup in $\Omega_{\rm b}$ seen in Fig.~1b at $t\sim 2-2.2$~Gyr and after
2.4~Gyr. The buckling instability also injects a (relatively) substantial disk
mass above its midplane of $z = \pm 1$~kpc for $r \ltorder r_{\rm VILR}$,
increasing the mass there by a factor of $\sim 6$, at $t\sim 2.4$~Gyr. We have
verified that particles which are injected above the plane are those trapped
by the bar prior and during the instability. Their specific angular momentum
is lower than for particles remaining in the plane by about 20\%. Lastly, the
ratio $(\sigma_{\rm z}/\sigma_{\rm r})^2$, of vertical-to-radial velocity
dispersions in the bar, drops during the bar growth to just below 0.4 and then
abruptly rises to about 0.95 at $t\sim 2.4$~Gyr. 

\section{Discussion}

The bar overall weakening and dissolution of the outer part during the buckling
instability are reflected in kinematical properties and changes of its orbital
structure. Those are discussed without invoking a specialized terminology.

The main kinematic change during the instability is that the bar which is
a very fast rotator initially, becomes a slow rotator between $\sim
2.1-3.3$~Gyr, and again a fast one afterwards. The definition `fast/slow' is
used here in the sense of the relative extent of the bar with respect to its
corotation (section~2). Because the characteristic offset dust lanes
delineating shocks exist only in a narrow range of a fast bar parameters
(e.g., Athanassoula 1992), we do not expect them to exist or, at least, to
have their usual shape, during $\sim 1$~Gyr of the buckling and some time
thereafter, when the bar is a slow rotator. In addition, the overall weakening
of the bar, which resembles more of an oval distortion during a prolonged
period of time, can affect the observed bar fraction, especially at higher
redshifts. This and subsequent bar growth are discussed elsewhere
(Martinez-Valpuesta et al., in preparation).

We have searched for main families of planar, $z=0$, and 3-D orbits which
support the bar shape before and after the buckling. Within the bar figure the
particles are largely trapped around prograde regular orbits which are aligned
with the bar. The bar has a narrow vertical extent prior to the buckling and
the particles, therefore, are largely confined to the bar midplane with two
degrees of freedom. In agreement with other studies, we find the 3-D prograde
families of orbits originating at the VILR and a retrograde family originating
from 1:1 resonance --- initially populated only very close to the bar $xy$
symmetry plane. During the buckling of the bar midplane, the main planar
orbits acquire the bent shape within $\sim r_{\rm VILR}$. The particles
injected above the plane (see section~2) are subsequently trapped on specific
3-D orbits (see section~1), whose $x$ and $z$ extensions increase sharply with
their energy. Populating them will provide the bar with the
peanut shape when viewed along its minor axis. The boxy shape appears somewhat
later (Fig.~2, lower panel), which has been also indicated by Raha et al.
(1991) for barred and by Patsis et al. (2002b) for nearly axisymmetric disks.  

An important observation by Friedli \& Pfenniger (1990) that the peanut/boxy
shapes appear even when the firehose instability in the bar is artificially
supressed by enforcing the symmetry in the $rz$ plane provides the crucial
insight into the role of this instability. The long evolutionary timescale of
the symmetrized bar is the result of the particle diffusion process enhanced
by the resonant heating via the VILR ---a secular process which ultimately
will furnish the bar with its characterisic peanut-shaped bulge, unless some
other more efficient heating of the stellar `fluid' will wash this out, e.g.,
star scattering by inhomogeneous gas (Berentzen et al. 1998). 
 
How does bar buckling change this picture? The important point here appears to
be particle injection above the disk plane (section~2) --- those populate
the characteristic 3-D family of orbits on a dynamical timescale, not
secularly. Thus, the buckling accelerates the process by breaking the symmetry
and by taking the evolution from the 2nd-order diffusion process to the
1st-order dynamical instability. 

What exactly is responsible for the sharp drop in bar's strength and size? The
orbital structure of the bar does not itself provide the answer, as it does
not supply us directly with the `population census' of different orbit
families. However, some measure of this, especially the fraction of phase
space occupied by regular orbits, is given by the surfaces of sections (e.g.,
Binney \& Tremaine 1987). The bar strength is growing steadily prior to
instability, and its ellipticity in the outer part is high and peaked at $\sim
0.8$. Strong bars are known to generate chaos, starting from near the
corotation, and the stochastic region is expected to widen with the bar
strength (e.g., Contopoulos 1981). To test this, the representative surfaces
of sections corresponding to regions deep inside the bar, at
intermediate radii and at the bar ends are shown in Fig.~3. At $t=1.6$~Gyr,
the bar interior is dominated by a trapped regular orbits, prograde for $y>0$
(the left two frames). The outer bar is
dominated by stochastic orbits and the invariant curves have dissolved here
(upper right frame). At $t=2.4$~Gyr, however, while the stochastic region is
still visible around the end of the bar, at $x=5.9$~kpc, its fraction has
decreased substantially --- in tandem with the decreased bar strength. Note,
that while the bar corotation propagates outwards, the stochastic region
expands inwards into the bar, dissolving its part outside the peanut shape
which is built at the VILR, and
weakening the bar further inside. The phase space regularity is largely
restored after the potential is mostly symmetrized again in the $rz$ plane.
Escaping chaotic orbits from the dissolving region have axial ratios by far
different from what is needed to support the bar and rapidly precess out of
the `valley' of the bar potential --- triggering a runaway dynamical process,
concurrent with the buckling instability.  
Fig.~4 shows one of the highest energy stable main planar orbits in the bar at
$t=1.6$~Gyr and, for the comparison, the similar energy orbit after the
buckling --- much shorter in its $x$ extention, less oval and confined to
within $r_{\rm VILR}$. This demonstrates the dominant trend in the bar
evolution during the instability. It also underscores that the outer bar
dissolution is concurrent with the buckling and maybe affected by it. The bar
cannot dissolve completely as this process is clearly limited to $r\gtorder
r_{\rm VILR}$.

\begin{figure}[ht!]
\vbox to1.7in{\rule{0pt}{1.7in}}
\includegraphics{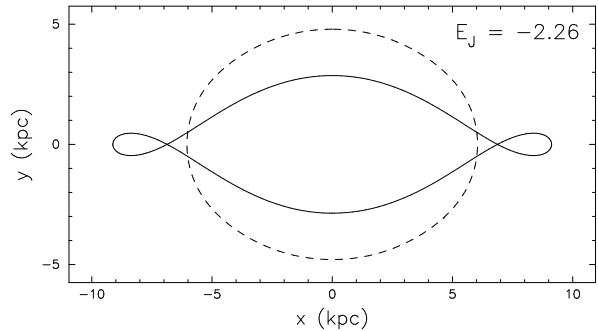}
\caption{Evolution of the main family of periodic orbits supporting the bar
figure at the same Jacobi energy, $E_{\rm J}=-2.26$, corresponding to a long
axis of $x\sim 9$~kpc at $t=1.6$~Gyr (solid) and $x\sim 6$~kpc at $t=2.4$~Gyr
(dashed). 
\label{fig:mass}
}
\end{figure}

In summary, a young stellar bar weakens overall and its outer part beyond 
the VILR dissolves after the bar reaches its peak strength. We show that this 
happens due to the inward expansion of the stochastic region near corotation. 
This effect is triggered by an exceptional strength of the growing bar prior 
to buckling instability. The rapidly developing chaos leads to a
self-destruction of the outer bar. There is no indication that the inner part
of the bar within the VILR can dissolve. The corollary is that
the bar becomes a slow rotator for about $\sim 1$~Gyr, a prolonged
period of time, with observational consequences for gas dynamics within the bar
and the formation of characteristic dust lanes there. The sharp decrease
in the bar strength for a similar period of time can have implications for
the observed bar fractions in local and higher redshift Universe, when both
spontaneous and tidally-induced bars are important.

\acknowledgments
We are indebted to Ingo Berentzen, Amr El-Zant, Clayton Heller and Barbara
Pichardo for fruitful discussions and comments on the manuscript.
I.S. is supported by NASA grants NAG 5-10823 and 5-13063 and NSF AST-0206251.
Additional partial support was provided by NASA through AR-09546 and 10284,
from STScI which is operated the AURA, Inc., under NASA contract NAS5-26555.
I.M. acknowledges partial support from PPARC. Simulations and orbital analysis
have been performed on a dedicated Linux cluster and we thank Brian Doyle for
technical support.

{}

\end{document}